\documentclass[fleqn,usenatbib]{mnras}
\usepackage{newtxtext,newtxmath}
\usepackage[T1]{fontenc}
\DeclareRobustCommand{\VAN}[3]{#2}
\let\VANthebibliography\thebibliography
\def\thebibliography{\DeclareRobustCommand{\VAN}[3]{##3}\VANthebibliography}

\usepackage{graphicx}	
\usepackage{amsmath}	

\title[Host galaxies of unlocalized FRBs in SDSS IV]{Search for host galaxies of unlocalized Fast Radio Bursts in the SDSS IV catalog}

\author[L.\'A. Garc\'ia et al.]{
Luz \'Angela Garc\'ia$^{1}$\thanks{E-mail: lgarciap@ecci.edu.co},
Eduard Piratova-Moreno$^{2}$,
Felipe Gonz\'alez-Alarc\'on $^{1}$,
and Jhonier Rangel$^{1}$
\\
$^{1}$ Universidad ECCI, Cra. 19 No. 49-20, Bogot\'a, Colombia, C\'odigo Postal 111311\\
$^{2}$ Fundaci\'on Universitaria Los Libertadores, Cra. 16 No. 63A-68, Bogot\'a, Colombia, C\'odigo Postal 111221
}

\date{Accepted XXX. Received YYY; in original form ZZZ}

\pubyear{\the\year{}}

\begin{document}
\label{firstpage}
\pagerange{\pageref{firstpage}--\pageref{lastpage}}
\maketitle

\begin{abstract}
This theoretical work investigates different models to predict the redshift of Fast Radio Bursts (FRBs) from their observed dispersion measure (DM) and other reported properties. We performed an extensive revision of the FRBs with confirmed galaxy hosts in the literature and compiled an updated catalog. With this sample of FRBs, composed of 117 unique transients, we explore four physically motivated models that relate the DM and redshift ($z$): a linear trend (inspired by the Macquart relation), a log-parabolic function, a power-law, and a combined model from the above. We assess the success of these theoretical proposals by implementing different statistical metrics and ranking them. The DM-$z$ relations are also tested using 100 realizations of 500 simulated FRBs, which follow the observed DM trends. Relying on our theoretical modeling, we establish the probability of $\sim$1000 FRBs with unknown $z$ (from the latest CHIME data release) to be hosted by galaxies in the SDSS archival dataset. Our validation scheme allows us to predict the FRBs with a probability threshold of $\geq$0.95 to originate in these galaxies, using their 2D angular position in the sky, magnitude in the r-band, and redshift. This statistical proposal will be tested with upcoming data releases from DESI and new generations of galaxy surveys, such as Euclid, and it opens brilliant possibilities to localize these transients in an automatic pipeline.
\end{abstract}


\section{Introduction}
One of the most challenging topics in modern astronomy is understanding the origin, nature, and evolution of Fast Radio Bursts (FRBs), a unique class of extragalactic transients. Since their discovery by \cite{lorimer2007}, thousands of FRBs have been detected by various radio telescopes around the world, within a wavelength range of 100 MHz to 8 GHz. These radio signals are observed randomly across the sky, with typical brightnesses reaching tens of Jy and durations of only a few milliseconds \citep{ma_gao2025}. Moreover, FRBs exhibit an apparent bimodality in their population: it remains unclear whether all of these radio bursts eventually repeat or not. If they do, repeating FRBs could originate from different progenitors than non-repeating ones \citep{beniamini2025}.\newline

The main observable used to characterize FRBs is the dispersion measure (DM), which quantifies the time delay between low- and high-frequency signals as photons from the FRB travel to the telescope. The total dispersion measure is typically expressed as the sum of three main contributions: i) the scattering due to our own Galaxy (MW); ii) a term associated with the host galaxy (host), often including a contribution from its halo; and iii) the dispersion due to inter- and circum-galactic medium (IGM and CGM, respectively) evolve with redshift ($z$), density, and temperature, etc.\newline
In general, the dispersion measure can be written as:
\begin{equation}
    \text{DM}=\text{DM}_{\text{MW}}+ \frac{\text{DM}_{\text{host}}}{1+z} + \text{DM}_{\text{IGM}}.
\end{equation}
The latter term, DM$_{\text{IGM}}$, can serve as a tracer of cosmological parameters—for example, to estimate the present-day value of the Hubble parameter \citep{kalika2024,konar2024,acharya2025,xu_feng2025,piratova2025,cvn2025}—or to study the large-scale structure of the Universe through the spatial distribution of FRBs \citep{hussaini2025}.\newline

Fast Radio Bursts are typically detected by radio telescopes, with a growing census now numbering a few thousand transients. However, radio antennas have limited angular resolution and cover large regions of the sky, creating a technical challenge: the redshift of these transients is generally unmeasured, resulting in an incomplete characterization of FRBs. For this reason, extensive efforts have been made to localize FRBs by identifying their host galaxies. On one hand, arcsecond-level localization of transients using interferometers has enabled the direct identification of their hosts. Furthermore, optical follow-up observations at the FRB position have enabled astronomers to identify optical counterparts coinciding with the radio source, often leading to the discovery of the \textit{true} host galaxy. However, this technique frequently yields faint or low-luminosity host candidates, introducing potential observational biases that are difficult to overcome in real-time observations \citep{james2025}. \newline
An illustrative example of the challenges involved in identifying host candidates for individual FRBs occurred in 2025 with FRB 20250316A. Multiple telescopes operating across different wavelengths reported possible host galaxies, with counterpart alerts ranging from the optical and infrared to the UV and X-ray bands. Interestingly, the unprecedented localization of this FRB—associated with the spiral galaxy NGC 4141—was made possible thanks to the record-breaking luminosity of the transient, which remains the brightest FRB detected to date. However, it is important to note that this is not the case for most FRBs, which are routinely observed with much lower luminosities. Therefore, identifying the host galaxies of FRBs with an electromagnetic follow-up campaign is computationally expensive, time-consuming, and biased towards bright FRBs, usually the least common. \newline

On the other hand, \citet{aggarwal2021,hamner2025} and subsequent works have proposed and applied a Bayesian framework to identify host-galaxy candidates for FRBs with unknown redshift ($z$): PATH (Probabilistic Association of Transients to their Hosts). This method incorporates priors on galaxy magnitude (typically in the r-band), the angular offset between the FRB position and the galaxy center, and the probability that the true host is not detected in the image. As a result, this observationally based approach provides a posterior estimate for each galaxy candidate’s probability of being the true host, as well as the probability that the real host remains undetected.\newline
The method relies on the strong correlation found between the apparent magnitude of already-identified host galaxies and the dispersion measure (DM) of their associated FRBs -assuming the Macquart relation \citep{macquare2020}-. This correlation arises from the fact that the apparent magnitude serves as an indicator of the galaxy’s distance to the observer, while DM$_{\text{IGM}}$ ultimately traces the cosmic distance between the FRB and the observer.\newline

Although programs such as PATH incorporate sophisticated algorithms to correct for biases and uncertainties in galaxy properties (particularly when derived from imaging surveys), any method aiming to localize the hosts of short transients must contend with positional offsets of several arcseconds on the sky for distant and extremely short-lived events, as well as with the high density of galaxies in the field. These factors introduce a non-negligible probability that an FRB may overlap with multiple galaxy sources.\newline

Finally, a completely different approach can be considered in rare cases: when a gravitational-wave detection occurs simultaneously with the burst of a radio transient. In such cases, the joint GW–FRB association could be used to infer the redshift of the latter from the luminosity distance measured for the former, as explored by \citet{qiang2024}.\newline

Alternatively, we propose a pipeline inspired by the search for electromagnetic counterparts to transients, which combines classical statistical methods, numerical algorithms, and our refined and updated DM–$z$ models \citep{piratova2024,piratova2025}. We build a database of 117 FRBs with confirmed redshifts to validate and test a novel method that automates the search for host-galaxy candidates of FRBs with unknown $z$. Taking advantage of the spectroscopic (secured) redshifts of archival galaxies in SDSS (Sloan Digital Sky Survey) -IV, we implement a complete end-to-end pipeline to compute the probability that galaxies in this spectroscopic survey host FRBs whose redshifts remain unknown due to the absence of optical or multi-wavelength follow-up observations.\newline
We use SDSS as a benchmark for this technique because it provides both spectroscopic and photometric information for galaxies, the program has been completed, and the dataset is fully available. These characteristics allow us to perform an efficient online search for galaxies around the two-dimensional position of each FRB of interest. In the future, an improved version of this schematic method can be extended to other spectroscopic surveys.\newline

This paper is organized as follows. Section~\ref{sec:metho} presents our proposed pipeline, including our DM–$z$ relations, the validation scheme using existing localized FRBs that overlap with the SDSS-IV galaxy catalog, the data augmentation method used to compute the relative probabilities for each model, and the metrics applied to rank our theoretical DM–$z$ functions. Section~\ref{sec:best_fits} reports the best-fit parameters for our sample of 117 localized FRBs, while Section~\ref{sec:first_results} describes the testing of our methodology with the confirmed FRB sample. Furthermore, Section~\ref{sec:final_prob} presents the results obtained when applying our approach to unlocalized FRBs from the latest CHIME/FRB data release that coincide with the SDSS footprint. We identify the most likely host-galaxy candidate for each FRB with a probability above 0.95. Finally, Section~\ref{sec:conclusions} summarizes the main findings of this study and discusses the limitations to be addressed in future work. Throughout this paper, we assume a flat $\Lambda$CDM cosmology with parameter estimates from \cite{planck2018}.

\section{Methodology}\label{sec:metho}

The methodology used to compute the probability that a galaxy in the SDSS catalog hosts a given FRB involves several steps. The complete pipeline described in this section was tested and calibrated using well-localized FRBs compiled by \citet{piratova2025}, as well as new confirmed FRBs reported by \citet{chimecol2025}. In total, we compile a catalog of 117 FRBs with the following characteristics in Table~\ref{tab:latest_frb_confirmed} (see Appendix D): right ascension ($ra_F$), declination ($dec_F$), dispersion measure and error (DM, $\Delta$DM), and redshift and its corresponding error -if reported- ($z$, $\Delta z$). In \citet{piratova2025}, we also included an additional feature: whether the FRB is a repeater (or not). However, the repeating nature of the transients is irrelevant to constrain their exact localization -to our current knowledge \cite{pastor2025,liu2025}-; thus, hereafter, it is no longer mentioned in our analysis. All the analysis is run in the platform \textsc{Deepnote} \citep{deepnote}\footnote{\url{https://deepnote.com/}}.
We transform the angular coordinates with the library \textsc{Astropy}\footnote{\url{https://www.astropy.org/}} to facilitate the search for galaxies in the SDSS online archive. \newline
Five transients from our analysis are excluded from the initial sample of 117 confirmed FRBs: ``FRB20220509G'', ``FRB20220914A'' \citep{connor2023}, and ``FRB20200120E'' \citep{zhang2024}, since their confirmed redshift is associated with a galaxy cluster (not a galaxy itself); ``FRB20221027A'' due to its poor host localization \citep{gao2024} and ``FRB20220529'', with loose constraints on its redshift. Including these FRBs will make our DM-$z$ relations quite unreliable.\newline
Our final sample of 112 FRBs with confirmed redshift ($z$) allows us to explore different functional forms that relate each FRB's observed dispersion measure (DM) and its $z$. In addition to the relations discussed in \citet{piratova2024}, we introduce an additional model to describe the individual terms of the total DM. 

\subsection{Modeling of the DM-$z$}\label{models}

Based on the DM-$z$ relations first explored in \citet{piratova2024}, we propose an additional model for the dispersion measure as a function of $z$ of the 112 transients in our analysis:\newline
\begin{table*}
\centering
\caption{Different DM-$z$ relations implemented in our analysis.} \label{tab:models}
\begin{tabular}{|c|l|l|}
\hline
\textbf{Model}  & & \textbf{Functional form}   \\ \hline
A & Linear model           & DM$=a \cdot z + b$                       \\
B & Log-parabolic function & DM$=$10$^{(a\cdot (\text{log }z)^2 + b\cdot \text{log }z + c - \text{log }0.3)}$ \\
C & Power-law function     & DM$=a(1+z)^{\alpha} + \text{DM}{_\text{MW}}$       \\
D & Combined model        & DM$=$10$^{(a\cdot (\text{log }z)^2 + b\cdot \text{log }z + c)} + d(1+z)^{\alpha}  +\text{DM}{_\text{MW}}$\\
\hline                  
\end{tabular}
\end{table*}

Here, the coefficients and exponents in Table~\ref{tab:models} are free parameters of each model, and their best fits will be presented in the next section, followed by a complete statistical analysis.\newline
The linear model was first introduced by \citet{macquare2020} and has been extensively applied in the literature, with new confirmed FRBs reported by \citet{cui2022,baptista2023,piratova2024}. This assumption works well, particularly at low $z$, when the cosmological distances grow linearly with $z$, almost unaffected by the underlying cosmic model.\newline
On the other hand, the logarithmic parabolic function was independently proposed and numerically modeled by \citet{pol2019} and \citet{zhu2021} to precisely describe the contribution of the intergalactic medium to the total dispersion measure. Using hydrodynamical simulations, they found that this empirical formula for DM-$z$ works effectively at low $z$ regimes when IGM plays a minor role in the complete dispersion of the FRB's photons. Following numerical findings by \citet{pol2019,zhu2021}, we assume that the contribution of IGM is approximately 30\% of the total DM at low $z$, which is reflected in the factor log (0.3) that appears in that case in Table~\ref{tab:models}. \newline
The power-law function explored by \citet{zhang2020,wang2023} presents a relation designed to model the DM caused by the host galaxy of the transient and the dispersion of our galaxy (DM$_{\text{MW}}$). The former term in this third model makes a significant contribution to the observed DM. In contrast, the term associated with the Milky Way can be modeled differently but consistently falls within the 40-100 pc/cm$^3$ range. Our analysis assumes that DM$_{\text{MW}} =$ 50 pc/cm$^3$.\newline
Finally, the latest (or combined model) in Table~\ref{tab:models} accounts for the fact that the dispersion measure can be described as one term due to our galaxy DM${_\text{MW}}$, a second dispersion caused by the FRB's host galaxy -numerically explained by a power-law function-, and a third term that describes the dispersion that occurs in the IGM.\newline
However, our goal is to use these DM-$z$ relations for FRBs with observed DM (and unknown $z$). Therefore, we require relations of the form $z= z(\text{DM})$, which means that we need to invert the functions presented in Table~\ref{tab:models}. In most cases, inversion is not straightforward, and interpolation is necessary. 

\subsection{Validation with localized FRBs}\label{sec:validation}
To localize and match the observed FRBs in our data set with previously detected and characterized galaxies, we use a spectroscopic galaxy catalog that includes a well-measured (secured) $z$, specifically, the latest SDSS-IV catalog\footnote{\url{https://cas.sdss.org/dr18/SearchTools/sql}}. \newline
We have previously checked and identified the catalog(s) in which the confirmed host galaxy has been observed, so we know in advance the number of confirmed FRBs in which the galaxy host is in the SDSS archival data. From the initial sample of 112 FRBs, only 23 coincide with the SDSS footprint. \newline
We implemented a self-consistent and automatic search for archival objects in the SDSS database -web scraping with the package \citep{bsref}- around the 2D coordinates of each confirmed FRB ($ra_{\text{F}}, dec_{\text{F}}$). We count all objects inside a 0.15 deg radius and retrieve the following features for all of them: object \texttt{ID}, \texttt{$ra_{\text{g}}$}, \texttt{$dec_{\text{g}}$}, apparent magnitudes in the \textit{u}, \textit{g}, \textit{r}, \textit{i} and \textit{z} filters (and their corresponding errors), \texttt{$z_{\text{g}}$}, \texttt{$\Delta z_{\text{g}}$}, \texttt{$z_{\text{warn}}$} and the class of object. From this list of objects, we remove those that belong to the class \texttt{STAR} and store only \texttt{GALAXY} or \texttt{QSO}. \newline
The angular separation $\theta_i$ (which measures the 2D angular distance between the FRB and each of the $N$ galaxies) is given by:

\begin{equation}
    \theta_{i} = \text{arccos}\left[\text{sin}(dec_{\text{F}})\text{sin}(dec_{\text{g; i}}) 
    + \text{cos}(dec_{\text{F}})\text{cos}(dec_{\text{g; i}})\text{cos}(ra_{\text{F}}-ra_{\text{g; i}})\right],
\end{equation}

\noindent with $ra_\text{F}, ra_{\text{g}}, dec_\text{F}, dec_{\text{g}}$ in radians, and $\sigma_{\theta} =$ 1 arcsec and $\sigma_{z} =$ 0.01, according to the uncertainties reported for spectroscopic surveys.\newline
The probability of a galaxy $i$ in the SDSS catalog (from $N$ objects inside the 0.15-degree circumference centered on the coordinates of the FRB) to host an FRB is given by $P_i$.
\begin{equation}\label{prob_sdss}
    P_{i} = \text{exp}\left(-\frac{\theta_{i}^2}{2\sigma_{\theta}^2}\right) \cdot \text{exp}\left(-\frac{(z_{\text{F}}-z_{\text{g; i}})^2}{2\sigma_{z}^2}\right).
\end{equation}

Since the three spatial coordinates of the FRBs are known and confirmed (for our validation dataset), we can rely on the definition of the probability in eq.~\eqref{prob_sdss} to calculate the probability of unknown FRBs to be hosted by galaxies in the SDSS repository. The largest source of uncertainty for transients without a secure $z$ is, in fact, the prediction of the redshift from the dispersion measure. Thus, our next stage in this pipeline is to rank our DM-$z$ models and assess their performance based on the predictions for our 112 well-localized bursts.

\subsection{Data augmentation for the validation dataset}\label{sec:pipeline}

Given that we have a reduced sample of confirmed FRBs (112 in total), we design a statistical scheme to evaluate the stability and precision of the proposed models with limited data, using the technique of \textit{bootstrap} \citep{chernick2012} that generates subsamples through random sampling with replacement and with different proportions of the original sample. The \textit{bootstrap} method enables us to introduce some randomness to the sample of observed FRBs, increase the amount of data available for evaluation and model fitting, and mitigate (or at least reduce) data biases. There are intrinsic observational biases that we need to account for, such as the Malmquist bias, which likely leads us to detect the brightest or closest FRBs, potentially affecting our conclusions with a small sample of known transients. \newline

The bootstrap estimate of the expectation value, which provides an unbiased estimate of the parameter’s expectation value under resampling  of a given variable $\theta$ (in our case, the free parameters of the models A-D), is: 
\begin{equation}
\hat{\theta}^{*}= \frac{1}{B}\sum_{b=1}^{B} \hat{\theta}_{b}^{*},
\end{equation}
\noindent where $B$ is the total bootstrap resamples and $\hat{\theta}_{b}^{*}$, the value of the $b$-th bootstrap sample. The variance of the bootstrap estimate is given by:
\begin{equation}
\hat{\sigma}^{2}_{{\theta}^{*}}= \frac{1}{B-1}\sum_{b=1}^{B} ( \hat{\theta}_{b}^{*}- \bar{\theta}^{*})^2 \:\:\:\:\: \text{with} \:\:\:\:\: \bar{\theta}^{*} = \frac{1}{B}\sum_{b=1}^{B} \hat{\theta}_{b}^{*}.
\end{equation}

For this study, we consider five different sampling proportions (0.75, 0.8, 0.85, 0.9, 0.95) to explore the variability of the performance of our models under distinct data configurations. For each model and subset proportion, we generate $b =$ 5,000 random samples (using different seeds, defined by the iteration index) to ensure the variability of the input data. This strategy ensures that every execution has a representative sample and generates a robust fit for each model.
For each random sample, we fit our models $z = z(\text{DM})$  with optimization techniques, according to the distribution of their residual errors, such as Ordinary Least Squares (OLS), Trust Region Reflective (\texttt{TRR}), Levenberg-Marquardt (\texttt{LM}), and \texttt{dogbox}. After trying the performance of all the packages, the one with the better response to the inversion of the DM-$z$ relations is \texttt{TRR} \citep{manguri2023}; thus, we use it hereafter.\newline

To find the best-fit values for each of the 5,000 realizations for each sub-sample proportion, using the four functional forms that relate DM to $z$ (as shown in Table~\ref{tab:models}), we incorporate different optimization techniques into our pipeline with \textsc{Scipy}. Moreover, we assess the performance of the theoretical relations through the following statistical metrics: the mean square root error (MSE), the likelihood $\mathcal{L}$, Bayesian Information Criteria (BIC), and the adjusted determination coefficient $R^2$. The following metrics are implemented to prevent overfitting and ensure a good fit for future datasets.
\begin{equation}\label{mse}
\text{MSE} = \sum_{i=1}^{n}(z_{obs; i}  -  z_{mod; i})^2.
\end{equation}
\noindent where $z_{mod; i}$ and $z_{obs; i}$ are the predicted and observed redshifts of the FRBs in our sample. The likelihood function is defined as follows:
\begin{equation}\label{like}
\text{ln} \mathcal{L} = -\frac{n}{2}\left[\text{ln}(2\pi) + 1 + \text{ln} \left( \frac{\text{MSE}}{n} \right)\right],   
\end{equation}
\noindent with $n$ the number of data points in our sample, that is, 112 localized FRBs. The variance of the maximum likelihood estimator is given by:
\begin{equation}
 \hat{\sigma}^2 = \frac{\text{MSE}}{n}.  
\end{equation}

The BIC is given by:
\begin{equation}\label{bic}
\text{BIC} = -2\text{ln}\mathcal{L} + K\cdot \text{ln}(n).   
\end{equation}
\noindent Here, $K$ is the number of free parameters of each model. The adjusted determination coefficient, $R^2_{\text{adj}}$ is given by the expression:
\begin{equation}\label{R2}
R^2_{\text{adj}} = 1 - \frac{(1 - R^2)(n - 1)}{n - p - 1},    
\end{equation}
\noindent with $R^2$ is the determination coefficient of the sample (see the full expression below) and $p$, the number of independent variables (in our case, $p$ corresponds to 1):
\begin{equation*}
R^2 = 1 - \frac{\sum_i (z_{obs; i} - z_{mod; i})^2}{\sum_i (z_{obs; i} - \bar{z})^2},    
\end{equation*}
\noindent here, $\bar{z}$ is the mean of the replicates. \newline

Once we have computed the best parameters and their corresponding errors for all subsamples, we select the subsampling proportion that best fits the observational data with 0.95 of the augmented dataset. Finally, we use the DM-$z$ relations with the best-fit parameters to compute the probability that a given FRB (with unknown $z$) will be hosted by a galaxy in SDSS, as explained by eq.~\eqref{prob_sdss}, but this time, the redshift is not confirmed as in the validation data set, but instead we have to predict it with our theoretical models.\newline
We only display the results for the best galaxy to host the FRB if the probability is above a 0.95 threshold and if the redshift of the FRB and the galaxy differ by less than 0.1.\newline

Finally, we implement a ranking system in our DM-$z$ models to determine which can provide the most reliable results in localizing FRBs when their redshift has not been confirmed with an electromagnetic counterpart. We apply three metrics to rank our models:
\begin{enumerate}
\item The maximum likelihood function calculated with the best-fit parameters of each model.
\item An adjusted geometrical probability, based on the geometrical probability discussed in \citet{aharonyan2018}.
\item The widely Applicable Information Criterion (WAIC).
\end{enumerate}

The adjusted geometrical probability satisfies the following criteria: i) a positive number in the range of [0,1]; ii) a quantity that measures precision and accuracy with respect to the observed data; iii) when a model's coverage is larger, its corresponding calculated probability is lower; iv) the predicted probability for each model cannot depend on the number of data points. \newline
To determine the precision of each model, we account for the area covered by the error bands predicted by the models ($A_{\text{mod}}$) and compare it with the area occupied by the observed data ($A_{\text{ref}}$), that stays invariable along the treatment, such that:
\begin{equation}
P_{\text{geo; pre}} = \text{exp}\left(-0.5 \cdot \frac{A_{\text{mod}}}{A_{\text{ref}}} \right),
\end{equation}
\noindent with $A_{\text{ref}} = (z_{obs; max} - z_{obs; min}) \cdot (\text{DM}_{obs; max} - \text{DM}_{obs; min})$. Now, the accuracy is defined here using the determination coefficient in the following way: 
\begin{equation}
P_{\text{geo; acc}} = \text{exp}\left(-0.5 \cdot (1 - R^2) \right),
\end{equation}
\noindent where the factor $1 - R^2$ allows us to warranty that if $R^2 \rightarrow$ 1, so does the probability. Interestingly, \citet{chicco2021} demonstrated that the determination coefficient can be interpreted as the inverse of a metric. The factor $(R^{2})^k$, with $k$ a constant value, is a true metric defined in the topological sense. In our case, $k$-value is assumed as 1.\newline
Finally, by multiplying both probabilities computed based on the precision and accuracy of each model, we find a probability in each case:
\begin{equation}\label{p_geo}
P_{\text{geo}} = P_{\text{geo; pre}} \cdot P_{\text{geo; acc}}
\end{equation}
On the other hand, the Widely Applicable Information Criterion is calculated as follows \citep{goldstein2022}:
\begin{equation}\label{waic}
\text{WAIC} = -2 \lambda_j + 2 p_{waic, j}. 
\end{equation}
\noindent WAIC metric penalizes the likelihood from a Bayesian point of view, even without the explicit use of a specific prior. Even in frequentist contexts, when a non-informative prior distribution is assumed, the WAIC can be applied, since this evaluation method does not incorporate essential previous knowledge of the system.\newline
The first and second terms on the right side of eq.~\eqref{waic} are given by:
\begin{equation}
\lambda_j = \sum_{i=1}^{n} \text{log}(P(z_i \vert N_i)) + \text{log}(P(N_i \vert \mu_i)),  
\end{equation}
\begin{equation}
p_{waic, j} = \text{var}\left(\sum_{i=1}^{n} \text{log}(P(z_i \vert N_i)) + \text{log}(P(N_i \vert \mu_i))\right),  
\end{equation}
\noindent where \( P(z_i|N_i) \) is the likelihood function or conditional distribution of the observed \( z_i \) given the set of model parameters \( N_i \), that is, the probability of observing the i-th FRB redshift under the fitted model. \( P(N_i|\mu_i) \) is a non-informative prior distribution of the model parameters centered at their mean value \( \mu_i \), reflecting the absence of strong prior knowledge and allowing the inference to rely mainly on the observed data. Finally, var denotes the variance of the logarithms of these probabilities, which measures the dispersion of the statistical evidence across the candidate models.\newline

With this outcome, we can assess which model responds better to the observed data and predict those FRBs that could be hosted by archival galaxies in the SDSS catalog.\newline

It is worth noting that there is some evidence suggesting that FRBs exhibit diversity at different redshift regimes. Thus, we repeat the entire pipeline for confirmed FRBs at $z \leq$0.5 in Appendix A (95 transients of our sample of localized FRBs in Table~\ref{tab:latest_frb_confirmed}). Furthermore, we create synthetic FRB catalogs to determine the robustness of our results with 100 mocks of 500 FRBs that follow a Weibull distribution fitted with the data from our catalog \citep{gupta2001}, which accounts for the evolution of the DM with $z$ in Appendix B. We run this mock test with localized FRBs at $z \leq$0.5 in Appendix C. 

\section{Best-fits for our DM-$z$ models based on confirmed FRBs}\label{sec:best_fits}

Here, we present an updated version of the DM-$z$ relations discussed in Section~\ref{models} with our data set of 112 localized FRBs. 
\begin{itemize}
\item Linear model (model A):
    \begin{equation}
    a = 971.68^{+162.23}_{-160.13}\: \text{pc}/\text{cm}^3, \:\:\:\:\: b = 226.56^{+46.78}_{-45.16} \: \text{pc}/\text{cm}^3.
\end{equation}
\item Log-parabolic function (model B):
\begin{equation}
    a = 0.20^{+0.13}_{-0.12} \: \text{pc}/\text{cm}^3, \:\:\:\:\:  b = 0.77^{+0.21}_{-0.23} \: \text{pc}/\text{cm}^3, \:\:\:\:\: c = 3.07^{+0.07}_{-0.09} \: \text{pc}/\text{cm}^3.
\end{equation} 
\item Power-law function (model C): 
\begin{equation}
    a = 240.26^{+36.59}_{-39.31} \: \text{pc}/\text{cm}^3, \:\:\:\:\:  \alpha = 3.28^{+0.51}_{-0.34},
\:\:\:\:\: \text{DM}_{\text{MW}} = 50 \: \text{pc}/\text{cm}^3.
\end{equation}
\item Combined model (model D):
\begin{equation*}
    a = -0.22^{+0.39}_{-5.50} \: \text{pc}/\text{cm}^3, \:\:\:\:\:  b = 0.02^{+1.20}_{-7.26} \: \text{pc}/\text{cm}^3, \:\:\:\:\: c = 2.53^{+0.47}_{-0.01} \: \text{pc}/\text{cm}^3, 
\end{equation*} 
\begin{equation}
    d = 99.44^{+100.57}_{-59.44} \: \text{pc}/\text{cm}^3, \:\:\:\:\:  \alpha = 3.88^{+2.08}_{-2.38}.
\end{equation} 
\end{itemize}

\begin{figure*}
\centering
\includegraphics[scale=0.32]{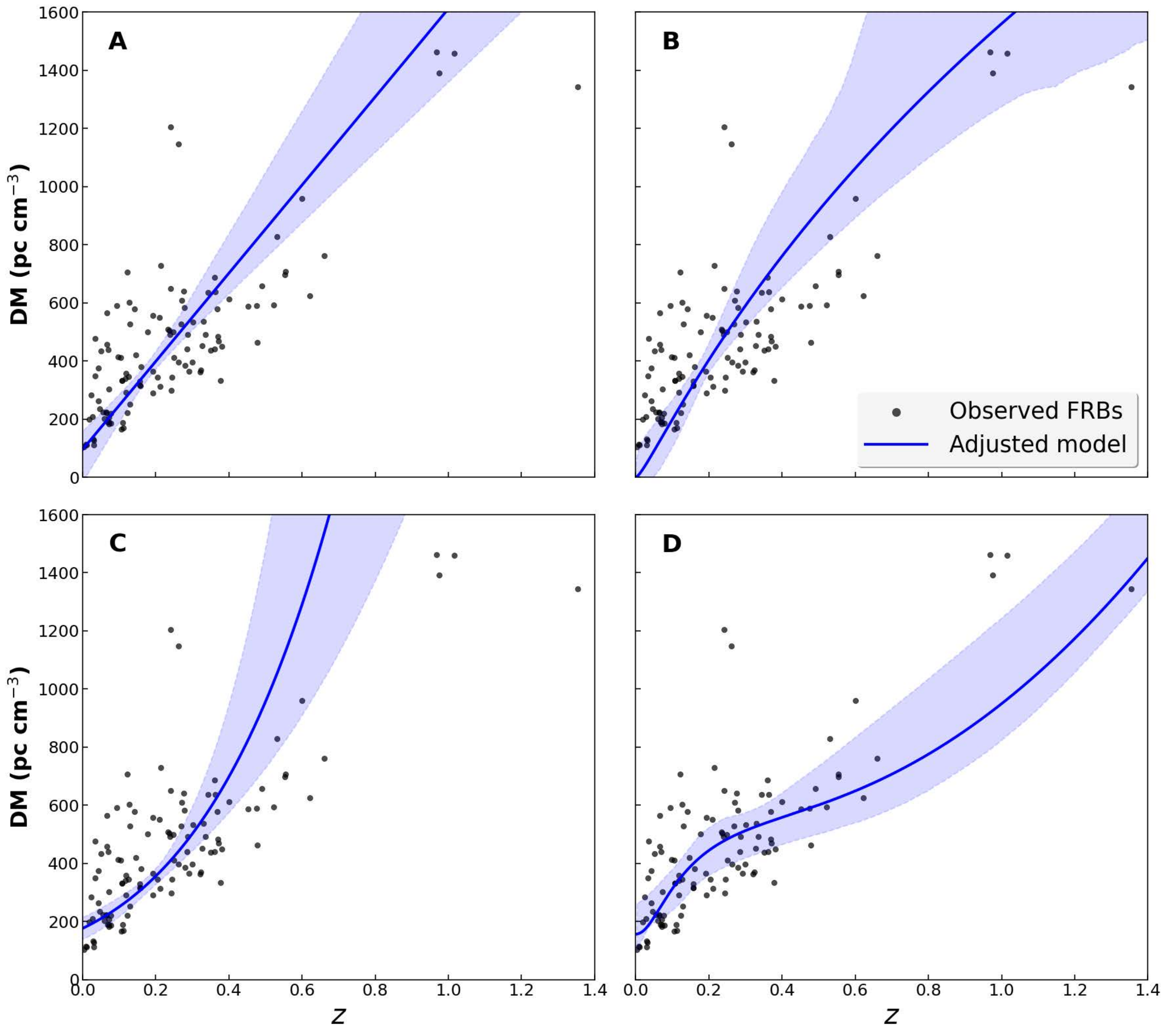}
\caption{DM-$z$ relations best-fits adjusted and compared with our 112 confirmed FRBs. Panels A-D show the prediction of each model: blue solid lines show the result with the best-fit parameters displayed above, and the shadowed lighter regions present the error bands for the linear trend (A), log-parabolic function (B), power-law (C), and combined model (D).}
\label{fig:dm-z-112-obs}
\end{figure*}

With the best parameters for all the models reported above, we plot our DM-$z$ relations in Figure~\ref{fig:dm-z-112-obs}.

\section{Validation of FRBs in SDSS-IV}\label{sec:first_results}
As a result of the web scraping implemented in our localized FRBs dataset, we found 23 transients that matched the footprint of the SDSS galaxy survey. \newline

Table~\ref{tab:validation} presents the main properties of each FRB and the galaxy most likely to be its host in SDSS. We count only galaxies within a 0.15-degree radius search around a given radio transient.\newline

\begin{table*}
\centering
\caption{FRBs and their most probable galaxy host in the SDSS galaxy survey. Column 1: FRB ID. Columns 2 and 3: right ascension and declination of the transient. Column 4: redshift. Column 5: number of galaxies inside the circumference centered on the FRB's coordinates. Column 6: calculated probability based on our validation scheme. Columns 7 and 8: right ascension and declination of the most probable candidate. Columns 9 and 10: redshift of the galaxy (if reported, its corresponding uncertainty). Columns 11 and 12: apparent magnitude in the $r$ filter (and error).}\label{tab:validation} 
\resizebox{1.0\textwidth}{!}{
\begin{tabular}{|c c c c c c c c c c c c|}
\hline 
\textbf{FRB} & $ra_{\text{F}}$ & $dec_{\text{F}}$ & $z_{\text{F}}$ & \# gal & P$_{\text{F}}$ & $ra_{\text{g}}$ & $dec_{\text{g}}$ & $z_{\text{g}}$ & $\Delta z_{\text{g}}$ & $m_r$ & $\Delta m_r$ \\
& (deg) & (deg) & & & & (deg) & (deg) &  & & & \\
\hline 
20181223C &	180.9208&	27.5477 &	0.0302 &	27	& 1.0 &	180.9208 &	27.5477 &	0.0302 &	0.0	&	19.20 &	0.02 \\
20190110C &	249.2841 &	41.4823 &	0.1224 &	24	& 0.859 &	249.2841 &	41.4823 &	0.1681 &	0.0  &	19.11	& 0.02 \\
20190303A &	207.9966 &	48.1247 &	0.064 & 31 &	1.0	& 207.9966 & 48.1247 &	0.0644 &	0.0 	& 18.39	& 0.04 \\
20190425A &	255.6625 &	21.5767 &	0.0312 &	22 &	1.0 &	255.6625 &	21.5767 &	0.0312 & 0.0 &	18.04 &	0.01 \\
20190608B &	334.0204 &	-7.8989 &	0.1178 &	13 &	1.0 &	334.0204 &	-7.8989 &	0.1178 &	0.0 &	18.61 & 0.02 \\
20191106C &	199.5801 &	42.9997 &	0.1078 &	29 &	1.0 &	199.5801 &	42.9997 &	0.1078 &	0.0	& 	18.35 & 0.03 \\
20200430A & 229.6971 &	12.3038 &	0.1608 &	30 &	0.782 &	229.6971 &	12.3038	& 0.1106 & 0.0 & 	17.89 &	0.02 \\
20210603A &	10.2489 &	21.1421	& 0.1772 &	33 &	0.845 &	10.2489 &	21.1421 &	0.1769 &	0.0 & 18.57 &	0.02 \\
20211212A &	157.6821 &	1.6298 &	0.0715 &	14 &	0.946 &	157.6821 &	1.6298 &	0.0637 &	0.0	&	17.85 &	0.01 \\
20220105A &	209.0858 &	22.6163 &	0.2785 &	46 &	0.479 &	209.0858 &	22.6163 &	0.3221 &	0.0001 &	21.18	& 0.05 \\
20230203A & 151.6883 & 35.6751 & 0.1464	& 60	& 0.980 & 151.6883 & 35.6751 & 0.1414	& 0.0001 & 20.13 & 0.03\\
20230216A &	155.9384 &	1.4056 &	0.531 &	29 &	0.780 &	155.9384 &	1.4056 &	0.4784 &	0.0001 &	21.08	& 0.05 \\
20230222B &	238.7380 &	30.8999 &	0.11 &	21	& 1.0 &	        238.7380 & 30.8999 &	0.1099 &	0.0001	& 19.57	& 0.02 \\
20230703A	& 184.6191 &	48.7543 &	0.1184 &	37	& 0.987 & 184.6191	&   48.7543 &	0.1194 &	0.0001 &	19.22 &	0.016 \\
20231005A	& 246.0247 &	35.4965	& 0.0713	& 31 &	0.836	& 246.0246	&   35.4965 & 0.1220	& 0.0001 &	19.13 &	0.015 \\ 
20231128A	& 199.5801 &	42.9997 &	0.1079 &	30 &	0.999 &	199.58013 & 42.9997 &	0.1078	& 0.0001 &	18.38 &	0.04 \\
20231204A &	207.9966 &	48.1247	& 0.0644	& 30	& 0.998	&   207.9965	&   48.1247 &	0.0644 &	0.0001	& 18.39 &	0.04 \\
20231223C &	259.5444 &	29.4958	& 0.1059	& 27 &	1.0 &	    259.5444 &	29.4959 &	0.1059 & 	0.0001	& 18.84	& 0.02 \\
20231226A &	155.2117 &	6.1785 &	0.1569 &	26 &	0.769 &	155.2117 &	6.1785	& 0.1117 &	0.0 &	18.64 &	0.03 \\
20240114A &	322.1104 &	4.4574 &	0.13 &	32 &	0.948 &	    322.1104 &	4.4574 &	0.1246 &	0.0 &	17.86 &	0.02 \\
20240124A &	322.0217& 	4.3546 &	0.269 &	33 &	0.780 &	    322.0217 &	4.3546 &	0.2628 &	0.0 &	19.44	& 0.02 \\
20240201A &	149.8981 &	14.0982 &	0.0427 &	33 &	0.996 &	149.8981 &	14.0982 &	0.0431 &	0.0 &	17.59 &	0.01 \\
20240213A & 158.8221 &	9.0714 &	0.1185 &	19 &	0.805 &	158.8221 &	9.0714 &	0.1402 &	0.0 & 18.73 &	0.01 \\
\hline                  
\end{tabular}
}
\end{table*}

\begin{figure}
\centering
\includegraphics[scale=0.35]{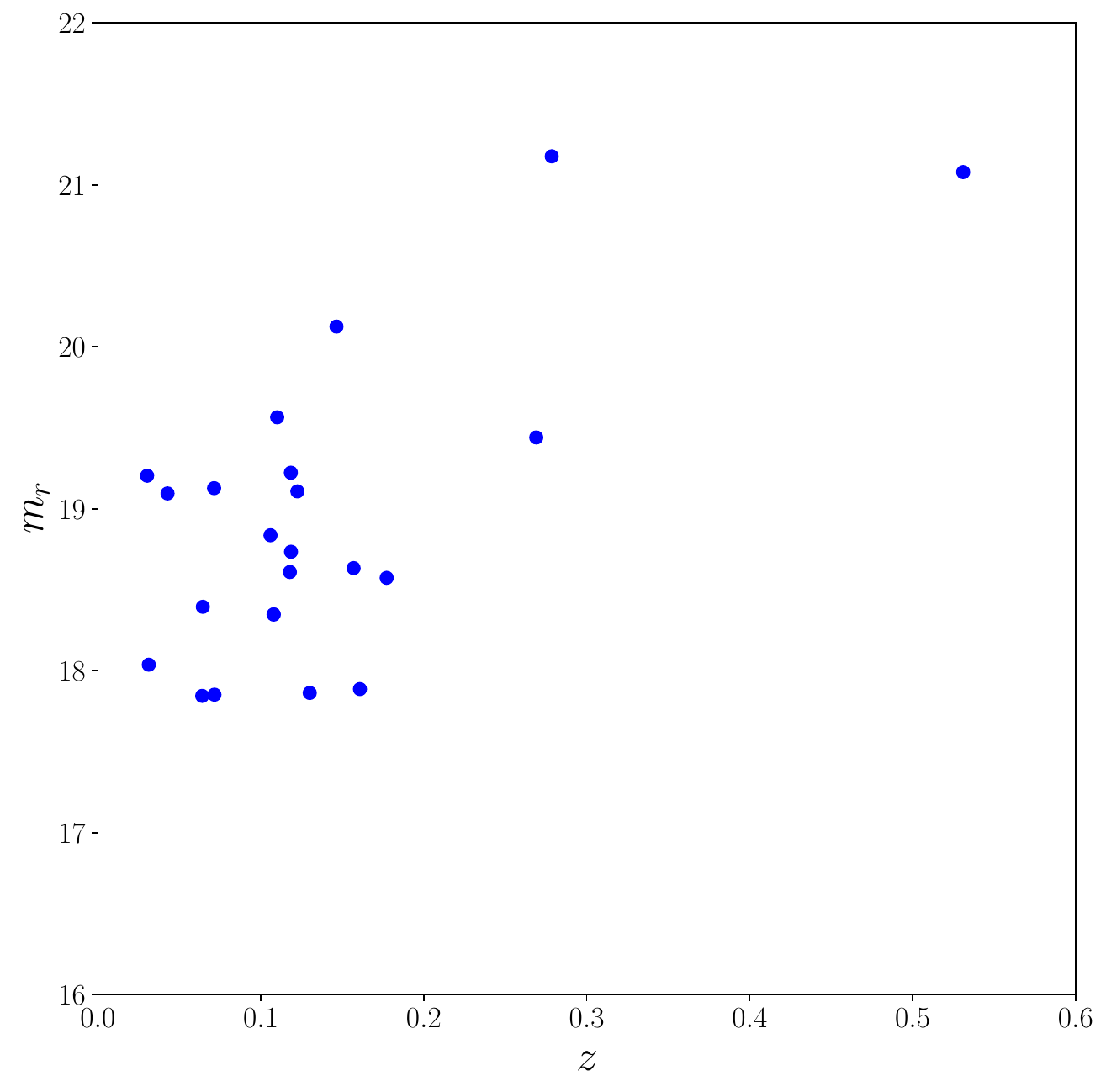}
\caption{Apparent magnitudes of the most likely transients' host galaxy as a function of the FRB's redshift. After running our validation scheme, we found 23 FRBs that match the SDSS footprint and presented them here.}
\label{fig:corr_val}
\end{figure}

As demonstrated by \citet{aggarwal2021} with a synthetic galaxy survey \textsc{GLADE}, there is a positive correlation between the apparent magnitude of the candidate host galaxy and the FRB dispersion measure. We compute this relation using the apparent magnitude in the \textit{r} filter\footnote{See top panel of Fig. 12 and Fig. 13 in \citet{aggarwal2021}.} of the most likely galaxy to host each one of the 23 FRBs from the validation set and the $z$ of the transients. The result is presented in Figure~\ref{fig:corr_val}.
Finally, we apply the three statistical metrics explained in subsection~\ref{sec:pipeline}, and find the results presented in Table~\ref{tab:metrics}.\newline
\begin{table}
\centering
\caption{Statistical metrics applied to rank our theoretical DM-$z$ relations from the validation dataset (FRBs with secured $z$).}\label{tab:metrics}
\begin{tabular}{|l|cccc|}
\hline
              & ln$\mathcal{L}$ & BIC     & $P_{\text{geo}}$ & WAIC \\
\hline              
Linear model     & - 685.83        & 1380.99 & 0.75             & 0.01 \\
Log-parabolic function & - 685.86        & 1385.72 & 0.74             & 0.99 \\
Power-law function    & - 687.15        & 1383.64 & 0.75             & 0.0  \\
Combined model     & - 681.59        & 1386.50 & 0.76             & 0.0 \\
\hline
\end{tabular}
\end{table}

It is worth noting that the best model needs to minimize the ln$\mathcal{L}$, exhibits the largest adjusted geometrical probability $P_{\text{geo}}$, as well as the highest value of the WAIC. We also included the computed BIC in Table~\ref{tab:metrics} for each model, to break a tie among them, given that the primary metrics considered give very close results. \newline  
Our findings with the statistical metrics implemented in our four models lead us to rank them in order of confidence as follows: i) Combined model (D); ii) Linear model (A); iii) Log-parabolic function (B); iv) Power-law function.

\section{Probability of FRBs with unknown $z$ to be hosted by a galaxy in the SDSS catalog}\label{sec:final_prob}

After performing a complete validation of our method, we are ready to test this pipeline with unlocalized radio transients. We extract 1022 FRBs from the publicly available data release TNS\footnote{\url{https://www.wis-tns.org/}} (unknown $z$) from \citet{chimecol2025} in March 2025. From these transients, only 912 have all the information required to run our pipeline ($ra_{\text{F}}$, $dec_{\text{F}}$, DM, and $\Delta$DM). From the latter subsample of FRBs, there are 270 that lie within the SDSS footprint. That is the set of selected FRBs that we have chosen to localize using the spectroscopic $z$ of this galaxy survey and the predicted $z$ from our models A-D. \newline

For each of the 270 FRBs, we run the pipeline described in subsection~\ref{sec:validation}, with each one of our theoretical DM-$z$ relations, leading to four sets of guesses of galaxies in the SDSS galaxy catalog for each FRB. We report the output of this method for the most likely galaxy host with a minimum threshold of 0.95 of the final probability, according to the model-ranking shown in Section~\ref{sec:first_results} in Tables~\ref{tab:res_bm}, ~\ref{tab:res_sec_bm}, ~\ref{tab:res_th_bm}, and ~\ref{tab:res_wm}. \newline

\begin{table*}
\centering
\caption{Most likely galaxy host of the unlocalized FRBs overlapping the SDSS galaxy survey with the Combined model (our best model according to the statistical metrics applied).}\label{tab:res_bm}
\resizebox{0.8\textwidth}{!}{
\begin{tabular}{|lcccccccc|}
\hline
\textbf{FRB} & $ra_{\text{g}}$ & $dec_{\text{g}}$ & $z_{\text{g}}$ & $ra_{\text{F}}$ & $dec_{\text{F}}$ & $z_{\text{F}}$ & P$_{\text{host}}$ & $\#$ gal\\
\hline
20180801A & 5.6230 & 1.2694 & 0.6061 & 5.6245 & 1.2692 & 0.6151 & 0.981 & 2 \\
20191021A & 2.1812 & 0.8103 & 0.1480 & 2.1811 & 0.8098 & 0.1478 & 0.986 & 156 \\
20241228A & 3.7778 & 0.2109 & 0.0682 & 3.7786 & 0.2107 & 0.0660 & 0.953 & 102 \\
20190531E & 0.2661 & 0.0098 & 0.1100 & 0.2660 & 0.0094 & 0.1092 & 0.987 & 414 \\
\hline
\end{tabular}
}
\end{table*}

\begin{table*}
\centering
\caption{Most likely galaxy host of the unlocalized FRBs overlapping the SDSS galaxy survey with the Linear trend (our second-best model according to the statistical metrics applied).}\label{tab:res_sec_bm}
\resizebox{0.8\textwidth}{!}{
\begin{tabular}{|lcccccccc|}
\hline
\textbf{FRB} & $ra_{\text{g}}$ & $dec_{\text{g}}$ & $z_{\text{g}}$ & $ra_{\text{F}}$ & $dec_{\text{F}}$ & $z_{\text{F}}$ & P$_{\text{host}}$ & $\#$ gal\\
\hline
20190226C & 0.3045 & 0.4664 & 0.4816 & 0.3053 & 0.4669 & 0.4824 & 0.952 & 113 \\
20190616A & 4.0848 & 0.6021 & 0.0767 & 4.0853 & 0.6020 & 0.0771 & 0.987 & 135 \\
20210530A & 5.8051 & 0.2849 & 0.6039 & 5.8050 & 0.2857 & 0.5970 & 0.958 & 76 \\
\hline
\end{tabular}
}
\end{table*}

\begin{table*}
\centering
\caption{Most likely galaxy host of the unlocalized FRBs overlapping the SDSS galaxy survey with the log-parabolic function (our third-best model according to the statistical metrics applied).}\label{tab:res_th_bm}
\resizebox{0.8\textwidth}{!}{
\begin{tabular}{|lcccccccc|}
\hline
\textbf{FRB} & $ra_{\text{g}}$ & $dec_{\text{g}}$ & $z_{\text{g}}$ & $ra_{\text{F}}$ & $dec_{\text{F}}$ & $z_{\text{F}}$ & P$_{\text{host}}$ & $\#$ gal\\
\hline
20190219B & 4.2990 & 1.0116 & 1.0903 & 4.2987 & 1.0111 & 1.1075 & 0.963 & 23 \\
20180909A & 2.1586 & 0.9905 & 0.1953 & 2.1576 & 0.9906 & 0.2021 & 0.972 & 100 \\
20211212A & 2.7512 & 0.0288 & 0.0961 & 2.7519 & 0.0293 & 0.1036 & 0.950 & 71 \\
20181017C & 5.7859 & -0.1546 & 0.1156 & 5.7854 & -0.1543 & 0.1196 & 0.978 & 43 \\
\hline
\end{tabular}
}
\end{table*}

\begin{table*}
\centering
\caption{Most likely galaxy host of the unlocalized FRBs overlapping the SDSS galaxy survey with the power-law function (our fourth-best model according to the statistical metrics applied).}\label{tab:res_wm}
\resizebox{0.8\textwidth}{!}{
\begin{tabular}{|lcccccccc|}
\hline
\textbf{FRB} & $ra_{\text{g}}$ & $dec_{\text{g}}$ & $z_{\text{g}}$ & $ra_{\text{F}}$ & $dec_{\text{F}}$ & $z_{\text{F}}$ & P$_{\text{host}}$ & $\#$ gal\\
\hline
20181224E & 4.1786 & 0.1284 & 0.3437 & 4.1783 & 0.1278 & 0.3441 & 0.963 & 69 \\
20190604D & 3.4902 & 0.2746 & 0.5258 & 3.4901 & 0.2745 & 0.5216 & 0.998 & 103 \\
20190414A & 3.1666 & 0.6796 & 0.4500 & 3.1671 & 0.6789 & 0.4478 & 0.963 & 127 \\
20191117A & 3.6771 & 0.8425 & 0.0709 & 3.6779 & 0.8421 & 0.0721 & 0.970 & 90 \\
20190317F & 4.1242 & 0.8209 & 0.5570 & 4.1239 & 0.8212 & 0.5541 & 0.993 & 89 \\
20190926B & 0.4440 & 0.5401 & 0.4995 & 0.4443 & 0.5409 & 0.5006 & 0.953 & 98 \\
\hline
\end{tabular}
}
\end{table*}

Our results offer a promising approach to localizing FRBs that are detected daily but lack confirmed redshifts, primarily due to the limited resolution of the radio telescopes that discover these transients and the absence of identified electromagnetic counterparts.\newline

So far, we have applied this methodology only to the SDSS, where all galaxies have spectroscopic redshifts and the dataset is fully available and systematized. However, the approach can be readily extended to other spectroscopic surveys and observatories.\newline

It is important to note that we report here only the galaxy from the SDSS-IV dataset with the highest probability of hosting each FRB. In most cases, multiple galaxies overlap with the FRB’s position within the survey area. Nevertheless, we focus exclusively on the galaxy that shows the smallest three-dimensional separation from the FRB coordinates ($ra_F$, $dec_F$, $z_F$), excluding galaxies that appear closer in angular position but whose predicted redshifts place them farther away than the FRB. This choice naturally introduces some uncertainty in our results, which we plan to address in future work.\newline

Finally, we emphasize the robustness of our methodology by comparing the best-fit parameters of our models with a subset of confirmed FRBs at $z \leq 0.5$ (see Appendix A). The parameters derived from both samples of confirmed FRBs are consistent, indicating that our main conclusions remain valid even if the FRB population exhibits a bimodal evolution at low and high redshifts ($z \leq 0.5$ and $z > 0.5$, respectively).\newline
In addition, the pipeline was tested and validated using 100 independent realizations of 500 synthetic FRBs each (Appendices B and C). This procedure not only improved the best-fit parameter values but, more importantly, significantly reduced their uncertainties across all models. Ultimately, synthetic samples containing five or more times the current number of observed FRBs will enable more robust predictions of FRB location at higher $z$, where the population’s distribution may evolve with cosmic time and thus may not be fully captured by the dispersion measure alone.

\section{Discussion and conclusions}\label{sec:conclusions}
We rely on the DM–$z$ relations previously explored in \citet{piratova2024} and \citet{piratova2025}, along with a classical statistical treatment, to develop a pipeline that estimates the probability of galaxy candidates hosting unlocalized FRBs. We test and validate our methodology using a dataset of 117 confirmed transients, and subsequently apply the pipeline to approximately 900 FRBs from the latest CHIME/FRB data release, for which no redshift information is available.\newline

Our probability estimates yield the most likely host galaxy for each of the 270 FRBs that overlap with the SDSS-IV archival galaxy catalog. This is achieved by measuring the angular offset between each FRB and the galaxies located within a 0.15° radius centered on the transient’s coordinates. Furthermore, our computed probabilities quantify the difference between the spectroscopic redshift of each galaxy and the predicted redshift for a given FRB, based on its reported DM and $\Delta$DM, according to our four theoretical DM–$z$ relations.\newline

Since only 23 of the 117 confirmed FRBs in our dataset fall within the SDSS footprint, we augment our sample by applying a bootstrap method to test the pipeline described in Section~\ref{sec:metho}. We generate 5,000 realizations to obtain the best-fit values of the free parameters in the DM–$z$ models. In addition, we produce 100 realizations of 500 mock FRBs to compare the results presented in Section~\ref{sec:best_fits} with synthetic data following a Weibull redshift distribution (see Appendix B).\newline

Using this framework and several statistical metrics—the maximum likelihood function, an adjusted geometrical probability, the Bayesian Information Criterion (BIC), and the Widely Applicable Information Criterion (WAIC)—we rank our DM–$z$ models by confidence level as follows: Combined model, Linear model, Log-parabolic function, and Power-law function. The most likely host galaxies for FRBs with unknown $z$ from the CHIME/FRB dataset are presented in Tables~\ref{tab:res_bm}, \ref{tab:res_sec_bm}, \ref{tab:res_th_bm}, and \ref{tab:res_wm}. Our most likely associations are summarized in Table~\ref{tab:unified_results}.
\begin{table}
\centering
\caption{Unified results from different models sorted by host probability (P$_{\text{host}}$) from highest to lowest.}\label{tab:unified_results}
\begin{tabular}{|lccc|}
\hline
\textbf{FRB} & \textbf{P$_{\text{host}}$} & \textbf{$\#$ gal} & \textbf{Model} \\
\hline
20190604D & 0.998 & 103 & Power-law \\
20190317F & 0.993 & 89 & Power-law \\
20190531E & 0.987 & 414 & Combined \\
20190616A & 0.987 & 135 & Linear \\
20191021A & 0.986 & 156 & Combined \\
20180801A & 0.981 & 2 & Combined \\
20181017C & 0.978 & 43 & Log-parabolic \\
20180909A & 0.972 & 100 & Log-parabolic \\
20191117A & 0.970 & 90 & Power-law \\
20190219B & 0.963 & 23 & Log-parabolic \\
20181224E & 0.963 & 69 & Power-law \\
20190414A & 0.963 & 127 & Power-law \\
20210530A & 0.958 & 76 & Linear  \\
20241228A & 0.953 & 102 & Combined \\
20190926B & 0.953 & 98 & Power-law \\
20190226C & 0.952 & 113 & Linear \\
20211212A & 0.950 & 71 & Log-parabolic \\
\hline
\end{tabular}
\end{table}

This schematic pipeline distinguishes itself from other programs in the field by focusing exclusively on localizing FRBs with unknown $z$ through our DM–$z$ relations and comparisons with spectroscopic archival data from SDSS-IV. Conversely, programs such as PATH \citep{aggarwal2021} rely primarily on photometric surveys and therefore do not use redshift information from the galaxy candidates. Instead, they rely on the apparent r-band magnitude, angular position, and other galaxy properties in the imaging data. Hence, this is the first pipeline that automatically searches for FRB host galaxies within a spectroscopic galaxy survey, using a framework that combines theoretical models, numerical algorithms, and classical statistical techniques to compute the probabilities.\newline

Interestingly, our methodology has so far been tested only on archival data from SDSS-IV, but it can be readily extended to other spectroscopic galaxy surveys in which the redshift of the galaxy is directly measured rather than inferred through indirect methods. In this sense, DESI and Euclid will be ideal for such applications, allowing our predictions to be tested over larger areas of the sky and in regions with a higher density of galaxies.\newline

Although our probabilities are currently computed using purely classical statistical methods, future work will focus on implementing a more rigorous Bayesian framework. While this is beyond the scope of the present study, it represents the natural next step for this program, as it will reduce the uncertainties associated with the redshift probabilities of FRBs.\newline

Furthermore, our proposal presents several caveats that must be addressed in future work. First, our method does not yet consider distinctive properties of the galaxy candidates beyond their angular position, spectroscopic redshift, and, in some cases, their apparent magnitude in the $r$-band. This simplification arises primarily from the lack of a clear correlation between the measured DM and the stellar mass ($M_s$), star formation rate (SFR), or metallicity of FRB host galaxies found in \citet{piratova2024}. The latter study was limited by a small sample of 22 confirmed FRBs, and therefore, those conclusions may not hold for larger datasets.\newline
More recent work by \citet{bernales2025} reports a strong correlation between DM$_\text{host}$ and both SFR and $M_s$. Similar trends have been identified by \citet{champati2025} and \citet{gupta2025}, whereas \citet{lei_wang2025} focused on the relationship between redshift and the cosmic SFR, reaching comparable conclusions. In another study, \citet{li2025} examined correlations between FRB and host-galaxy properties—including SFR, $M_s$, specific SFR (sSFR), inclination angle, and projected area—and found a moderate correlation only with the sSFR. Finally, \citet{glowacki2025} reported a strong correlation between FRB properties and the stellar surface density (compactness) of their host galaxies, as well as weak correlations with H$\alpha$ equivalent width and stellar gravitational potential, but none with host-galaxy inclination.\newline
All of these studies could inspire future extensions of our work by incorporating additional galaxy properties into the probability computation, ultimately improving the localization of FRBs with unknown $z$.\newline

This program will continue to evolve as more confirmed FRBs are added to our validation dataset. The main source of uncertainty at present arises from the predicted $z_{\text{FRB}}$, but this limitation will progressively decrease as the number of confirmed transients increases, as demonstrated in Appendices B and C through 100 random realizations of 500 synthetic FRBs in each mock with $z \sim 1.4$, as well as in our previous studies in this topic \citep{piratova2024,piratova2025}.

\section*{Conflict of Interest Statement}
The authors declare that the research was conducted in the absence of any commercial or financial relationships that could be construed as a potential conflict of interest.

\section*{Funding}
This work was supported by Fundaci\'on Universitaria Los Libertadores programme ``D\'ecimo Segunda (XII) Convocatoria Interna Anual de Proyectos de Investigaci\'on, Creaci\'on Art\'istica y Cultural", project ``Estimaci\'on del espacio de par\'ametros para modelos difusivos cosmol\'ogicos a trav\'es de m\'etodos bayesianos y de machine learning." [Grant number: ING-14-24].

\section*{Acknowledgments}
The authors thank Universidad ECCI and Fundaci\'on Universitaria Los Libertadores for granting us the resources to develop this project. This material is based upon work supported by the Google Cloud Research Credits program with the award GCP19980904. This research made use of \textsc{matplotlib} \citep{hunter2007}, \textsc{SciPy} \citep{virtanen2020}, \textsc{NumPy} \citep{harris2020}, \textsc{Pandas} \citep{pandasref} and \textsc{BeautifulSoup} \citep{bsref}.

\bibliographystyle{mnras}
\bibliography{mnras}

\appendix

\section{Appendix A. Best-fits for our DM-$z$ models based on confirmed FRBs at $z <$ 0.5} 
Here, we present an updated version of the DM-$z$ relations discussed in Section~\ref{models} with our data set of 112 localized FRBs. 
\begin{itemize}
\item Linear model (model A):
    \begin{equation}
    a = 847.40^{+220.33}_{-194.57}\: \text{pc}/\text{cm}^3, \:\:\:\:\: b = 248.14^{+49.42}_{-43.29} \: \text{pc}/\text{cm}^3.
\end{equation}
\item Log-parabolic function (model B):
\begin{equation}
    a = 0.01^{+0.06}_{-0.01} \: \text{pc}/\text{cm}^3, \:\:\:\:\:  b = 0.34^{+0.13}_{-0.08} \: \text{pc}/\text{cm}^3, \:\:\:\:\: c = 2.89 \pm 0.06 \: \text{pc}/\text{cm}^3.
\end{equation} 
\item Power-law function (model C): 
\begin{equation}
    a = 228.66^{+40.21}_{-34.56} \: \text{pc}/\text{cm}^3, \:\:\:\:\:  \alpha = 3.50^{+0.60}_{-0.55},
\:\:\:\:\: \text{DM}_{\text{MW}} = 50 \: \text{pc}/\text{cm}^3.
\end{equation}
\item Combined model (model D):
\begin{equation*}
    a = -0.89^{+0.92}_{-6.01} \: \text{pc}/\text{cm}^3, \:\:\:\:\:  b = -1.0^{+1.37}_{-7.17} \: \text{pc}/\text{cm}^3, \:\:\:\:\: c = 2.17^{+0.64}_{-0.01} \: \text{pc}/\text{cm}^3, 
\end{equation*} 
\begin{equation}
    d = 98.4^{+101.6}_{-58.4} \: \text{pc}/\text{cm}^3, \:\:\:\:\:  \alpha = 3.03^{+3.09}_{-1.53}.
\end{equation} 
\end{itemize}

\begin{figure*}
\centering
\includegraphics[scale=0.32]{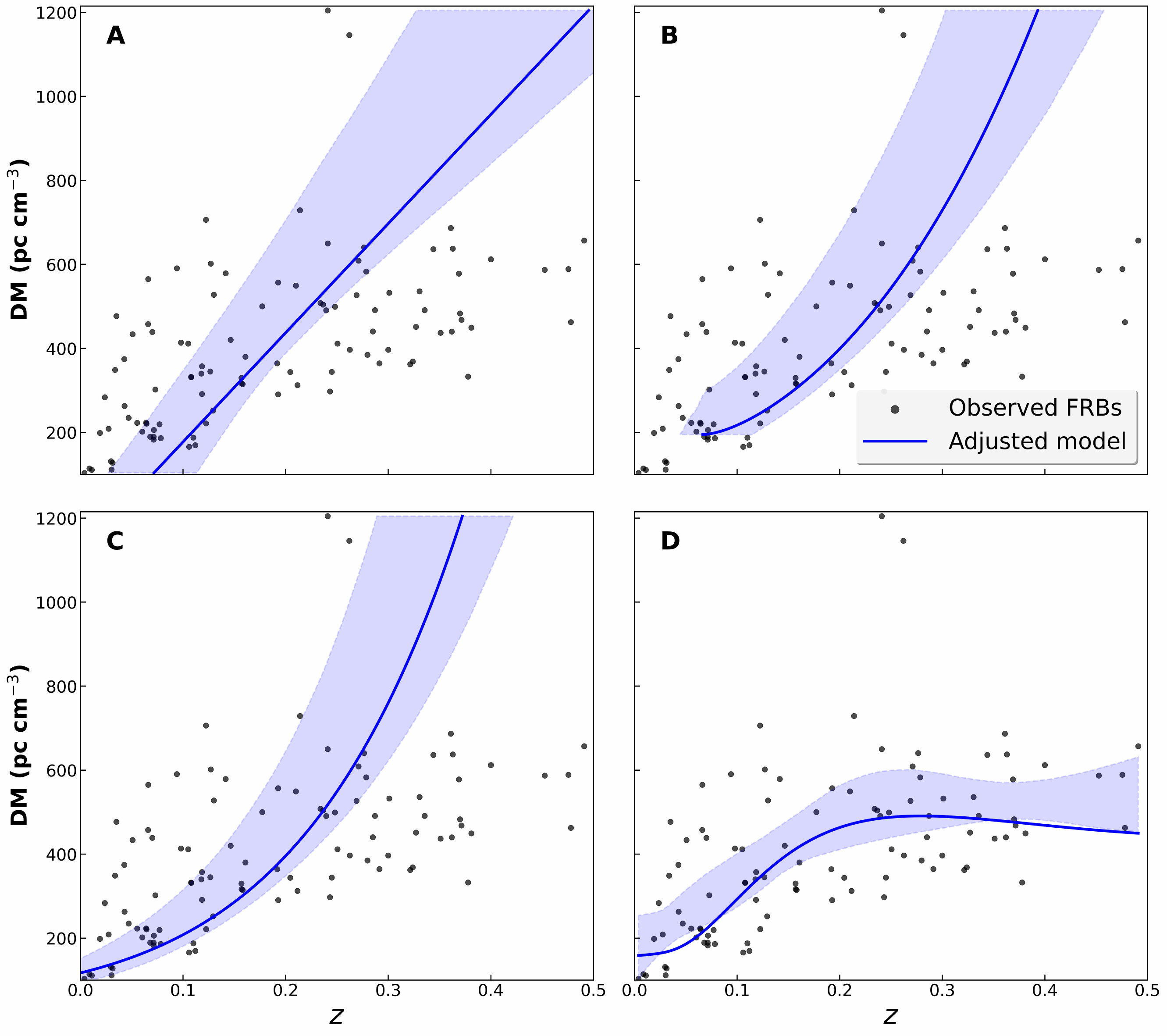}
\caption{DM-$z$ relations best-fits adjusted and compared with confirmed FRBs -with a redshift range of $z \leq$0.5 -. Panels A-D show the prediction of each model: blue solid lines show the result with the best-fit parameters displayed above, and the shadowed lighter regions present the error bands for the linear trend (A), log-parabolic function (B), power-law (C), and combined model (D).}
\label{fig:dm-z-95-obs}
\end{figure*}

With the best parameters for all the models reported above, we plot our DM-$z$ relations in Figure~\ref{fig:dm-z-95-obs}.

\section{Appendix B. Best-fits for our DM-$z$ models based on a synthetic sample of FRBs} 
This Appendix is devoted to presenting the best-fit parameters of the DM-$z$ relations discussed in Section~\ref{models} with 100 realizations of 500 synthetic FRBs. 

\begin{itemize}
\item Linear model (model A):
    \begin{equation}
    a = 1163.33^{+204.40}_{-160.35} \: \text{pc}/\text{cm}^3, \:\:\:\:\: b = -50.34^{+24.58}_{-39.12} \: \text{pc}/\text{cm}^3.
\end{equation}
\item Log-parabolic function (model B):
\begin{equation}
    a = 0.03^{+0.21}_{-0.15} \: \text{pc}/\text{cm}^3, \:\:\:\:\:  b = 0.89^{+0.25}_{-0.14} \: \text{pc}/\text{cm}^3, \:\:\:\:\: c = 3.07^{+0.07}_{-0.04} \: \text{pc}/\text{cm}^3.
\end{equation} 
\item Power-law function (model C): 
\begin{equation}
    a = 62.86^{+18.04}_{-10.80} \: \text{pc}/\text{cm}^3, \:\:\:\:\:  \alpha = 5.92^{+0.74}_{-0.87},
\:\:\:\:\: \text{DM}_{\text{MW}} = 50 \: \text{pc}/\text{cm}^3.
\end{equation}
\item Combined model (model D):
\begin{equation*}
    a = -0.50^{+0.47}_{-3.69} \: \text{pc}/\text{cm}^3, \:\:\:\:\:  b = 0.52^{+0.43}_{-2.56} \: \text{pc}/\text{cm}^3, \:\:\:\:\: c = 2.68^{+0.18}_{-0.86} \: \text{pc}/\text{cm}^3, 
\end{equation*} 
\begin{equation}
    d = 88.36^{+33.50}_{-28.60} \: \text{pc}/\text{cm}^3, \:\:\:\:\:  \alpha = 1.89^{+1.21}_{-0.92}.
\end{equation} 
\end{itemize}

\begin{figure*}
\centering
\includegraphics[scale=0.3]{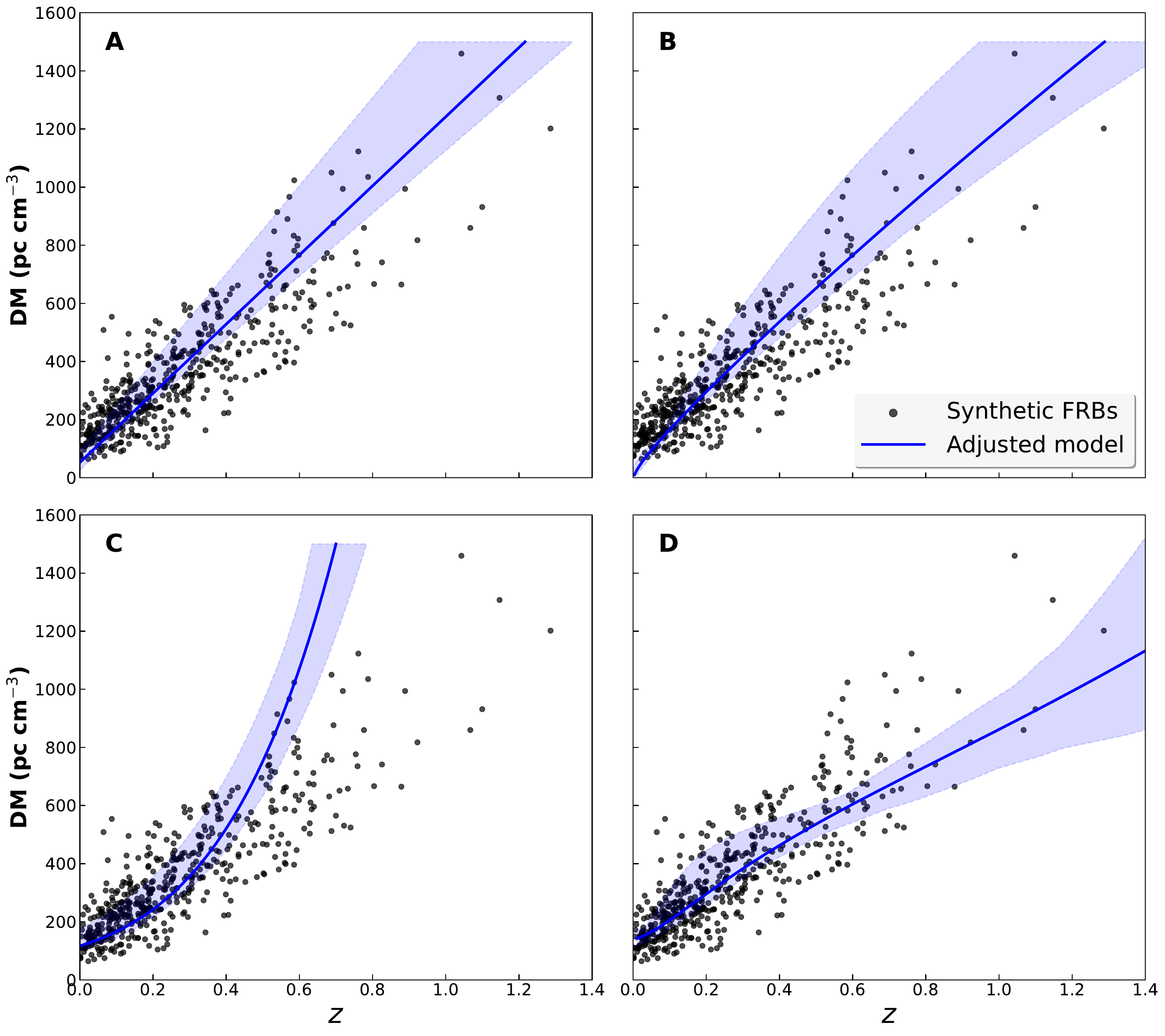}
\caption{DM-$z$ relations best-fits adjusted with 100 realizations of 500 synthetic FRBs, each. Panels A-D show the prediction of each model: blue solid lines show the result with the best-fit parameters displayed above, and the shadowed lighter regions present the error bands for the linear trend (A), log-parabolic function (B), power-law (C), and combined model (D).}
\label{fig:dm-z-112-syn}
\end{figure*}

With the best parameters for all the models reported above, we plot our DM-$z$ relations in Figure~\ref{fig:dm-z-112-syn}.

\section{Appendix C. Best-fits for our DM-$z$ models based on a synthetic sample of FRBs at $z <$ 0.5} 
Here, we present an updated version of the DM-$z$ relations discussed in Section~\ref{models} with 100 realizations of 500 synthetic FRBs. 

\begin{itemize}
\item Linear model (model A):
    \begin{equation}
    a = 1481.35^{+498.85}_{-148.02}\: \text{pc}/\text{cm}^3, \:\:\:\:\: b = -4.95^{+68.36}_{-30.19} \: \text{pc}/\text{cm}^3.
\end{equation}
\item Log-parabolic function (model B):
\begin{equation}
    a = 0.30^{+0.53}_{-0.26} \: \text{pc}/\text{cm}^3, \:\:\:\:\:  b = 1.44^{+0.92}_{-0.42} \: \text{pc}/\text{cm}^3, \:\:\:\:\: c = 3.33^{+0.40}_{-0.18} \: \text{pc}/\text{cm}^3.
\end{equation} 
\item Power-law function (model C): 
\begin{equation}
    a = 35.48^{+15.09}_{-10.59} \: \text{pc}/\text{cm}^3, \:\:\:\:\:  \alpha = 9.95^{+1.98}_{-0.69},
\:\:\:\:\: \text{DM}_{\text{MW}} = 50 \: \text{pc}/\text{cm}^3.
\end{equation}
\item Combined model (model D):
\begin{equation*}
    a = -1.89^{+1.46}_{-2.72} \: \text{pc}/\text{cm}^3, \:\:\:\:\:  b = -1.64^{+1.74}_{-2.73} \: \text{pc}/\text{cm}^3, \:\:\:\:\: c = 1.67^{+0.78}_{-1.17} \: \text{pc}/\text{cm}^3, 
\end{equation*} 
\begin{equation}
    d = 82.47^{+34.63}_{-29.14} \: \text{pc}/\text{cm}^3, \:\:\:\:\:  \alpha = 3.21^{+1.64}_{-1.70}.
\end{equation} 
\end{itemize}

\begin{figure*}
\centering
\includegraphics[scale=0.3]{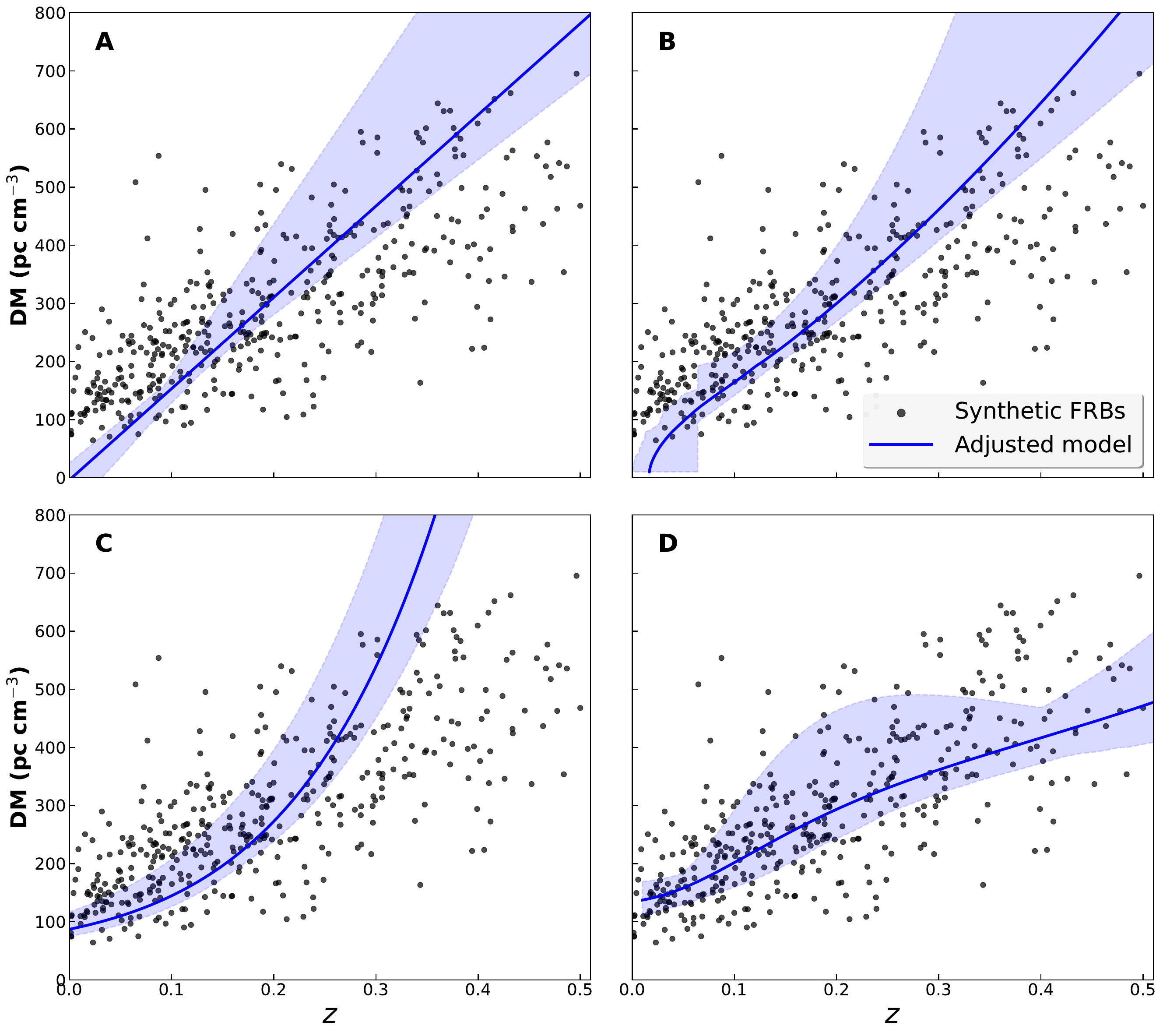}
\caption{DM-$z$ relations best-fits adjusted with 100 realizations of 500 synthetic FRBs, with a redshift range of $z \leq$0.5. Panels A-D show the prediction of each model: blue solid lines show the result with the best-fit parameters displayed above, and the shadowed lighter regions present the error bands for the linear trend (A), log-parabolic function (B), power-law (C), and combined model (D).}
\label{fig:dm-z-95-syn}
\end{figure*}

With the best parameters for all the models reported above, we plot our DM-$z$ relations in Figure~\ref{fig:dm-z-95-syn}.

\section{Appendix D. FRBs with confirmed $z$ to date}

\newpage
\begin{table*}
\centering
\caption{Confirmed FRBs to date. Column 1: FRB ID. Columns 2 and 3: right ascension and declination of the transient. Columns 4 and 5: observed dispersion measure and its uncertainty. Columns 6 and 7: redshift of the transient and error. FRBs indicated with a $\dagger$ are hosted by a galaxy cluster. Therefore, they are excluded from our localization pipeline. Similarly, FRBs marked with the $\star$ symbol are excluded from our analysis due to a poor identification of their galaxy host (thus, they do not have a secure $z$).}\label{tab:latest_frb_confirmed} 
\resizebox{0.6\textwidth}{!}{
\begin{tabular}{|l c c c c c c|}
\hline 
\textbf{FRB} & $ra_{\text{F}}$ & $dec_{\text{F}}$ & DM$_{\text{obs}}$ & $\Delta$DM$_{\text{obs}}$ & $z_{\text{F}}$ & $\Delta z_{\text{F}}$   \\
& (deg) & (deg) & (pc/cm$^3$) & (pc/cm$^3$) & & \\
\hline 
20121102A & 82.9946 & 33.1479 & 557.0000 & 2.0000 & 0.1927 & 0.0000 \\
20171020A & 333.7500 & -19.6667 & 114.1000 & 0.2000 & 0.0087 & 0.0000 \\
20180301A & 93.2292 & 4.6711 & 536.0000 & 5.0000 & 0.3305 & 0.0000 \\
20180814A & 65.6833 & 73.6644 & 189.4000 & 3.2300 & 0.0680 & 0.0000 \\
20180916B & 29.5031 & 65.7168 & 348.8000 & 1.6200 & 0.0337 & 0.0000 \\
20180924B & 326.1052 & -40.9000 & 362.1600 & 0.0600 & 0.3214 & 0.0000 \\
20181030A & 163.2000 & 73.7400 & 103.5000 & 1.6200 & 0.0039 & 0.0000 \\
20181112A & 327.3485 & -52.9709 & 589.0000 & 0.0300 & 0.4755 & 0.0000 \\
20181220A & 348.6982 & 48.3421 & 208.6600 & 1.6200 & 0.0275 & 0.0000 \\
20181223C & 180.9207 & 27.5476 & 111.6100 & 1.6200 & 0.0302 & 0.0000 \\
20190102C & 322.4157 & -79.4757 & 364.5450 & 0.3000 & 0.2913 & 0.0000 \\
20190110C & 249.3185 & 41.4434 & 221.6000 & 1.6200 & 0.1224 & 0.0000 \\
20190303A & 207.9958 & 48.1211 & 223.2000 & 1.6200 & 0.0640 & 0.0000 \\
20190418A & 65.8123 & 16.0738 & 182.7800 & 1.6200 & 0.0713 & 0.0000 \\
20190425A & 255.6625 & 21.5767 & 127.7800 & 1.6200 & 0.0312 & 0.0000 \\
20190520B & 240.5167 & -11.2883 & 1204.7000 & 4.0000 & 0.2410 & 0.0010 \\
20190523A & 207.0650 & 72.4697 & 760.8000 & 0.6000 & 0.6600 & 2.0000 \\
20190608B & 334.0199 & -7.8982 & 340.0500 & 0.5000 & 0.1178 & 0.0000 \\
20190611B & 320.7455 & -79.3976 & 332.6300 & 0.2000 & 0.3778 & 0.0000 \\
20190614D & 65.0755 & 73.7067 & 959.2000 & 0.5000 & 0.6000 & 0.0000 \\
20190711A & 329.4195 & -80.3580 & 592.6000 & 0.4000 & 0.5217 & 0.0000 \\
20190714A & 183.9797 & -13.0210 & 504.1300 & 2.0000 & 0.2365 & 0.0000 \\
20191001A & 323.0000 & -54.6667 & 507.9000 & 0.0400 & 0.2340 & 0.0000 \\
20191106C & 199.5801 & 42.9997 & 332.2000 & 0.0000 & 0.1078 & 0.0000 \\
20191228A & 344.4292 & -29.5942 & 297.5000 & 0.0500 & 0.2432 & 0.0000 \\
20200120E$^{(\dagger)}$ & 146.2500 & 68.7700 & 87.8200 & 1.6200 & 0.0008 & 0.0000 \\
20200223B  & 82.6950 & 288.3130 & 201.8000 & 0.0000 & 0.0602 & 0.0000 \\
20200430A & 229.7064 & 12.3769 & 380.1000 & 0.4000 & 0.1608 & 0.0000 \\
20200906A & 53.4958 & -14.0831 & 577.8000 & 0.2000 & 0.3688 & 0.0000 \\
20201123A & 263.6690 & -50.7672 & 433.5500 & 0.0000 & 0.0507 & 0.0000 \\
20201124A & 76.9900 & 26.1900 & 413.5200 & 3.2300 & 0.0979 & 0.0000 \\
20210117A & 339.9792 & -16.1517 & 728.9500 & 0.3600 & 0.2140 & 0.0010 \\
20210320C & 204.3200 & -15.4104 & 384.8000 & 0.3000 & 0.2797 & 0.0000 \\
20210405I & 255.3396 & -49.5452 & 565.1700 & 0.0000 & 0.0660 & 0.0000 \\
20210410D & 326.0862 & -79.3182 & 578.7800 & 0.0000 & 0.1415 & 0.0000 \\
20210603A & 10.2740 & 21.2260 & 500.1470 & 0.0040 & 0.1772 & 0.0001 \\
20210807D & 299.2042 & -0.8143 & 251.9000 & 0.2000 & 0.1293 & 0.0000 \\
20211127I & 199.7896 & -18.8246 & 234.8300 & 0.0800 & 0.0469 & 0.0000 \\
20211203C & 204.4700 & -31.3678 & 636.2000 & 0.4000 & 0.3439 & 0.0000 \\
20211212A & 157.6696 & 1.6769 & 206.0000 & 5.0000 & 0.0715 & 0.0000 \\
20220105A & 208.9642 & 22.4888 & 583.0000 & 1.0000 & 0.2785 & 0.0000 \\
20220204A & 278.3321 & 71.6157 & 612.2000 & 0.0000 & 0.4000 & 0.0000 \\
20220207C & 310.1995 & 72.8823 & 263.0000 & 0.0000 & 0.0430 & 0.0000 \\
20220208A & 319.3483 & 71.5400 & 437.0000 & 0.0000 & 0.3510 & 0.0000 \\
20220307B & 350.8745 & 72.1924 & 499.3280 & 0.0000 & 0.2480 & 0.0000 \\
20220310F & 134.7205 & 73.4908 & 462.6570 & 0.0000 & 0.4780 & 0.0000 \\
20220319D & 32.1779 & 71.0350 & 110.9500 & 0.0100 & 0.0110 & 0.0000 \\
20220330D & 165.7256 & 71.7535 & 468.1000 & 0.0000 & 0.3714 & 0.0000 \\
20220418A & 219.1056 & 70.0959 & 624.1240 & 0.0000 & 0.6220 & 0.0000 \\
20220501C  & 352.3792 & -32.4907 & 449.5000 & 0.2000 & 0.3810 & 0.0000 \\
20220506D & 318.0448 & 72.8273 & 396.6510 & 0.0000 & 0.3000 & 0.0000 \\
20220509G$^{(\dagger)}$ & 282.6700 & 70.2438 & 269.5300 & 10.0000 & 0.0894 & 0.0000 \\
20220529A$^{(\star)}$ & 19.1042 & 20.6325 & 246.0000 & 0.0000 & 0.1839 & 0.0000 \\
20220610A & 351.0000 & -33.5167 & 1458.1000 & 0.2000 & 1.0160 & 0.0020 \\
20220717A & 293.3042 & -19.2877 & 637.3400 & 0.0000 & 0.3630 & 0.0000 \\
20220725A & 336.7500 & 34.8833 & 290.4000 & 0.3000 & 0.1926 & 0.0000 \\
20220726A & 75.1058 & 71.6018 & 686.5500 & 0.0000 & 0.3610 & 0.0000 \\
20220825A & 311.9814 & 72.5850 & 649.8930 & 0.0000 & 0.2410 & 0.0000 \\
20220831A & 333.0854 & 71.5376 & 1146.2500 & 0.0000 & 0.2620 & 0.0000 \\
20220912A & 347.2704 & 48.7071 & 219.4600 & 0.0420 & 0.0771 & 0.0000 \\
20220914A$^{(\dagger)}$ & 282.0568 & 73.3369 & 631.2900 & 10.0000 & 0.1139 & 0.0000 \\
20220918A & 17.7412 & -70.7850 & 657.0000 & 0.4000 & 0.4910 & 0.0000 \\
\hline                  
\end{tabular}
}
\end{table*}

\begin{table*}
\centering
\caption{Table~\ref{tab:latest_frb_confirmed} (continued)}
\resizebox{0.6\textwidth}{!}{
\begin{tabular}{|l c c c c c c|}
\hline 
\textbf{FRB} & $ra_{\text{F}}$ & $dec_{\text{F}}$ & DM$_{\text{obs}}$ & $\Delta$DM$_{\text{obs}}$ & $z_{\text{F}}$ & $\Delta z_{\text{F}}$   \\
& (deg) & (deg) & (pc/cm$^3$) & (pc/cm$^3$) & & \\
\hline 
20220920A & 240.2571 & 70.9188 & 314.9770 & 0.0000 & 0.1580 & 0.0000 \\
20221012A & 280.7987 & 70.5242 & 440.3580 & 0.0000 & 0.2850 & 0.0000 \\
20221027A$^{(\star)}$ & 129.6104 & 71.7315 & 452.5000 & 0.0000 & 0.2290 & 0.0000 \\
20221029A & 143.8351 & 71.7529 & 1391.0500 & 0.0000 & 0.9750 & 0.0000 \\
20221101B & 341.4589 & 71.5295 & 490.7000 & 0.0000 & 0.2395 & 0.0000 \\
20221106A & 56.7048 & -25.5698 & 343.8000 & 0.8000 & 0.2044 & 0.0000 \\
20221113A & 72.8406 & 71.6131 & 411.4000 & 0.0000 & 0.2505 & 0.0000 \\
20221116A & 17.6617 & 71.5288 & 640.6000 & 0.0000 & 0.2764 & 0.0000 \\
20221219A & 255.7773 & 71.6817 & 706.7000 & 0.0000 & 0.5540 & 0.0000 \\
20230124A & 233.0768 & 71.7273 & 590.6000 & 0.0000 & 0.0940 & 0.0000 \\
20230203A & 151.6616 & 35.6941 & 420.1000 & 0.0000 & 0.1464 & 0.0000 \\
20230216A & 155.9717 & 1.4678 & 828.0000 & 0.0000 & 0.5310 & 0.0000 \\
20230222A & 106.9604 & 11.2245 & 706.1000 & 0.0000 & 0.1223 & 0.0000 \\
20230222B & 238.7391 & 30.8987 & 187.8000 & 0.0000 & 0.1100 & 0.0000 \\
20230307A & 177.7813 & 71.4100 & 608.9000 & 0.0000 & 0.2710 & 0.0000 \\
20230311A & 91.1097 & 55.9460 & 364.3000 & 0.0000 & 0.1918 & 0.0000 \\
20230501A & 338.5535 & 71.5292 & 532.5000 & 0.0000 & 0.3010 & 0.0000 \\
20230521B & 349.6785 & 71.5220 & 1342.9000 & 0.0000 & 1.3540 & 0.0000 \\
20230526A & 22.3646 & -52.7688 & 316.4000 & 0.2000 & 0.1570 & 0.0000 \\
20230626A & 240.7125 & 71.7142 & 451.2000 & 0.0000 & 0.3270 & 0.0000 \\
20230628A & 161.8999 & 71.7745 & 345.1500 & 0.0000 & 0.1265 & 0.0000 \\
20230703A & 184.6244 & 48.7299 & 291.3000 & 0.0000 & 0.1184 & 0.0000 \\
20230708A & 303.2371 & -55.3807 & 411.5000 & 0.0600 & 0.1050 & 0.0000 \\
20230712A & 170.7112 & 71.7794 & 586.9600 & 0.0000 & 0.4525 & 0.0000 \\
20230718A & 127.6129 & -41.0036 & 477.0000 & 0.5000 & 0.0350 & 0.0000 \\
20230730A & 54.6646 & 33.1593 & 312.5000 & 0.0000 & 0.2115 & 0.0000 \\
20230814A & 335.9746 & 73.0259 & 696.4000 & 0.0500 & 0.5535 & 0.0000 \\
20230902A & 52.3671 & -47.5626 & 440.1000 & 0.1000 & 0.3619 & 0.0000 \\
20230926A & 269.1249 & 41.8143 & 222.8000 & 0.0000 & 0.0553 & 0.0000 \\
20231005A & 246.0280 & 35.4487 & 189.4000 & 0.0000 & 0.0713 & 0.0000 \\
20231011A & 18.2411 & 41.7491 & 186.3000 & 0.0000 & 0.0783 & 0.0000 \\
20231017A & 346.7543 & 36.6527 & 344.2000 & 0.0000 & 0.2450 & 0.0000 \\
20231025B & 270.7881 & 63.9891 & 368.7000 & 0.0000 & 0.3238 & 0.0000 \\
20231120A & 143.6169 & 71.7574 & 438.9000 & 0.0000 & 0.0700 & 0.0000 \\
20231123A & 82.6232 & 4.4755 & 302.1000 & 0.0000 & 0.0729 & 0.0000 \\
20231123B & 240.5665 & 71.7156 & 396.7000 & 0.0000 & 0.2625 & 0.0000 \\
20231128A & 199.5782 & 42.9927 & 331.6000 & 0.0000 & 0.1079 & 0.0000 \\
20231201A & 54.5893 & 26.8177 & 169.4000 & 0.0000 & 0.1119 & 0.0000 \\
20231204A & 207.9990 & 48.1160 & 221.0000 & 0.0000 & 0.0644 & 0.0000 \\
20231206A & 112.4428 & 56.2563 & 457.7000 & 0.0000 & 0.0659 & 0.0000 \\
20231220A & 122.2054 & 71.7217 & 491.2000 & 0.0000 & 0.3355 & 0.0000 \\
20231223C & 259.5446 & 29.4979 & 165.8000 & 0.0000 & 0.1059 & 0.0000 \\
20231226A & 155.2817 & 6.1294 & 329.9000 & 0.1000 & 0.1569 & 0.0000 \\
20231229A & 26.4678 & 35.1129 & 198.5000 & 0.0000 & 0.0190 & 0.0000 \\
20231230A & 72.7976 & 2.3940 & 131.4000 & 0.0000 & 0.0298 & 0.0000 \\
20240114A & 322.0703 & 4.4841 & 527.7000 & 0.0000 & 0.1300 & 0.0000 \\
20240119A & 218.1169 & 71.7554 & 483.1000 & 0.0000 & 0.3700 & 0.0000 \\
20240123A & 66.1340 & 71.5965 & 1462.0000 & 0.0000 & 0.9680 & 0.0000 \\
20240124A & 321.9162 & 4.3501 & 526.9000 & 0.0000 & 0.2690 & 0.1390 \\
20240201A  & 149.9056 & 14.0880 & 374.5000 & 0.3000 & 0.0427 & 0.0000 \\
20240210A & 8.7796 & -28.2708 & 283.7300 & 0.0500 & 0.0237 & 0.0000 \\
20240213A & 158.7613 & 9.0000 & 357.4000 & 0.0000 & 0.1185 & 0.0000 \\
20240215A & 268.4333 & 71.6540 & 549.5000 & 0.0000 & 0.2100 & 0.0000 \\
20240229A & 173.7346 & 71.7838 & 491.1500 & 0.0000 & 0.2870 & 0.0000 \\
20240310A  & 17.6219 & -44.4394 & 601.8000 & 0.2000 & 0.1270 & 0.0000 \\
\hline                  
\end{tabular}
}
\end{table*}

\bsp	
\label{lastpage}
\end{document}